\documentstyle[proceedings]{iau177}

\def\stackg{\raisebox{-2.5pt}{$\stackrel{\textstyle{>}}{\sim}$} }

\begin{opening}

\title{Carbon Star Dust from Meteorites}

\author{ Uffe Gr{\AA}e J{\O}rgensen and Anja C.\ Andersen }

\institute{Niels Bohr Institute, Copenhagen University Observatory}
\end{opening}

\runningtitle{ Carbon Star Dust from Meteorites}

\begin{document}

\vspace*{-0.5cm} 

\begin{abstract}
Inside carbonaceous chondrite meteorites are tiny dust particles which, 
when heated, release noble gases with an isotopic composition different
from what is found anywhere else in the solar system. For this reason
it is believed that these grains are (inter)stellar dust which 
survived the collapse of the interstellar cloud that became the solar system.
We will describe here why we believe that the most abundant of these grains,
micro-diamonds, were formed in the atmospheres of carbon stars, and explain
how this theory can be tested observationally.

\end{abstract}

\section{Introduction}

The discovery of meteoritic dust grains with origin outside the solar 
system has opened
the possibility to study presolar material in the laboratory, with all 
the advantages in details and accuracy such analysis allow for.
Identification of the (possible) stellar origin of the meteoritic grains, 
offer us a unique opportunity to add important new
constrains on models of stellar evolution 
(detailed elemental and isotopic abundances) and stellar atmospheric
structure (elemental and mineralogical composition of the grains).
There are several indications that a fraction, possible the bulk, of the 
presolar meteoritic grains has its origin in carbon stars.

For the moment the amount of detailed information (like isotopic ratios of
tiny noble gas impurities) about the meteoritic grains is overwhelming
(see e.g.\ Zinner 1995), whereas much of the
fundamental data necessary in order to apply the meteoritic results to
stellar modelling is entirely missing. For example, the necessary rate 
coefficients for formation of the most abundant presolar meteoritic grains 
(diamond dust) are lacking, and stellar wind models therefore do not 
predict diamond formation, but instead such models predict amorphous carbon 
(which has not been identified in meteorites) as the most abundant grain 
type in carbon-rich environments. 

A combined self-consistent description of the full atmospheric
region of a red giant star does not exist yet, but is slowly becoming within 
reach. The meteoritic data combined with more fundamental laboratory data
can be an important ingredient in constructing such a model for the first time.
Successful construction of self-consistent models, followed by possibly 
verifications of the formation place(s) of the most abundant stellar grains
that contributed to the formation of the meteorites (and hence also the planets)
would provide us fundamental new knowledge about the sources of material for
the solar system and the chemical evolution of the Galaxy.

The most common types of meteorites are fragments of larger proto-planetary
bodies, which melted and chemically differentiated after their formation. 
Carbonaceous chondrites, on the other hand, are meteorites which have never 
been part of a larger body. They 
consist of spherical glass-like chondrules embedded in a fine-grained matrix.
The matrix has had a gentle thermal history and is believed to be 
the (relatively unprocessed) original dust from which the planets formed.
Therefore, the larger the amount of matrix is in the chondrite, relative 
to the chondrule material, 
the more of the original solar nebula material is present,
and the more primitive the chondrite is said to be.

When this matrix material was heated in the laboratory, it was
realized already in the early 1960'ies (see Lewis \& Anders 1983 for a 
review) that at certain temperatures the matrix released noble gases with 
an isotopic composition markedly different from everything else
in the solar system. It was therefore concluded that the matrix contains
one or more types of grains, formed before the solar system, in which
non-solar composition noble gases are trapped. 
After years of trials with different chemical purifications of the matrix 
material, and subsequent stepwise heating and isotopic noble gas measurements,
the first presolar grains were finally isolated by Lewis et al.\ in 1987, and 
identified as tiny diamonds.

Diamonds account for more than 99\% of the identified presolar meteoritic 
material, with an
abundance that can exceed 0.1\% (1000\,ppm) of the matrix (Huss \& Lewis 1994b),
corresponding to more than 3\% of the total amount of carbon in the meteorite.
The second and third most abundant types are SiC (6\,ppm) and graphite
(less than 1\,ppm). They are all chemically quite resistant, which makes it 
possible to isolate them
by dissolving the meteorite in acids. Further, a few of the SiC and graphite 
grains has been shown to contain tiny sub-grains of titanium and refractory 
carbides (Bernatowicz et al.\ 1991, 1992, 1994). Three isotopically anomalous,
non-carbon-bearing grains have also been found. They are corundum (Al$_2$O$_3$),
spinel (MgAl$_2$O$_4$), and silicon nitride (Si$_3$N$_4$)
(Russel et al.\ 1995, Nittler et al.\ 1994, 1995). In the following sections 
we will discuss diamonds, SiC and graphite in some detail.

\section{Diamonds}

The individual diamond grains are very small, with a
median diameter of less than 20\,\AA\ (Fraundorf et al.\ 1989).
Since the diamond lattice distance is about 2\,\AA, a typical presolar diamond
contains of the order $({20\over 2})^3$ = 1000 carbon atoms, 
with 6${\times}10^2$
$\approx$ 50\% of these belonging to the surface. Since surface atoms
have one unpaired bond, they will (in a hydrogen-rich atmosphere) resemble
hydrogenated amorphous carbon (a-C:H). Only the $\approx$50\% 
``interior'' atoms will sit in an actual diamond crystal structure. The
presolar diamonds are therefore often called amorphous diamonds.

It is not obvious to which degree the extracted diamonds resemble the original
diamond dust at its place of origin.
Many alternations could have occurred in interstellar space, in the solar 
nebula, as well as during the chemical extraction process in the laboratory.
However, the first step in an observational identification of their
astronomical source of origin might be modelling of their synthetic
spectrum. For this purpose we have measured the monochromatic absorption
coefficient, described which of the features can be
expected to be intrinsic to the diamonds (and which might be artifacts from
the chemical processing in the laboratory), and
computed synthetic carbon star spectra with the diamonds 
included (Andersen et al.\ 1996). The features which are most likely to
be intrinsic are listed in Table 1 and compared with the results 
obtained by other groups.

{\footnotesize
\begin{table}[h,t,b]
\caption[]{Spectral features, in cm$^{-1}$, detected in the spectra of the 
presolar diamonds from the Allende, Murchison and Orgueil meteorites.}
\begin{tabular}{lllllll}     \hline
\multicolumn{3}{c}{ALLENDE} & \multicolumn{2}{c}{MURCHISON} &
                ORGUEIL & ASSIGNMENT \\ \hline
(1)  &  (2)$^{a}$ &  (3) & (4) & (5) & (6)$^{a}$ &     \\
  &  & 50\,000 &   & 50\,000 &  & paired N in diamond \\
  &  & 37\,037   &   & 37\,037   &    & paired N in diamond \\
2919 &  & 2954  &  & 3000 &  2940 &     aliphatic C--H stretch  \\
2849 &  & 2854  &  & 2800 &       &         \\
1361 &  & 1385 & 1399 & & 1380 &     C--H deformation (CH$_{3}$) \\
 & & & & &   & /interstitial N \\
1173 & 1143 & 1122 & 1084 & 1175 & 1042 &  C--O/C--N stretch/ \\ 
1028 & 1089 & 1054 & & 1090 & & CH$_{2}$ waging \\
 &626 & & & & 620 & CH out-of-plane \\
& & & & 396, 367 & &  C=O=C or C=N=C \\
& & & & 310 & & \\
& & & & 130, 120 & & ?? \\ \hline
\end{tabular}
\footnotesize{(1)$\sim$Lewis et al.\,1989, (2)$\sim$Koike et al.\,1995,
(3)$\sim$Andersen et al.\,1996, (4)$\sim$Lewis 1992, 
(5)$\sim$Mutschke et al.\,1996, (6)$\sim$Wdowick et al.\,1988,
$a$ $\sim$ the spectra were obtained on diamond-like residues}
\end{table}
} %end footnotesize

Like for the other grains, the strongest argument that the diamonds are
formed outside the solar system is the peculiar, non-solar isotopic 
composition of their noble gas inclusions (and other trace element inclusions). 
There are several reasons why we believe the bulk of the diamonds form
in carbon stars, one of the most important ones being their $^{12}$C/$^{13}$C 
ratio of $\approx$ 90. This ratio is identical to what is observed
in carbon stars with large excess of carbon (i.e., with C/O \stackg 1.5, 
and strongly mass losing), and it is not found in any other
abundant astronomical objects.
In contrast to this, SiC (the second most abundant 
presolar grain) has $^{12}$C/$^{13}$C $\approx$ 40, which is 
typical (Lambert et al.\ 1986) for carbon stars with only small
excess of carbon (i.e., with C/O $\approx$ 1).
Hydrostatic {\sc marcs} photospheric models 
indicate that SiC grains will dominate the grain formation for
C/O $\approx$ 1 (where $^{12}$C/$^{13}$C is as found in the meteoritic SiC 
grains), whereas pure carbon grains will dominate for the high
C/O ratios (where $^{12}$C/$^{13}$C is as in the meteoritic diamonds).
This was the primary basis for our theory (J{\o}rgensen 1988)
that diamonds come from
evolved carbon stars and SiC from less evolved carbon stars (actually,
at the time the paper was written it was a prediction that SiC should 
exist in meteorites). The (radiative pressure driven) mass loss increases
rapidly with increasing C/O (= increasing $^{12}$C/$^{13}$C) of the stars.
If diamonds and SiC are formed in carbon stars, it is therefore a natural 
consequence of this theory that
the meteorites contain much more diamonds than SiC. A more quantitative
simulation is still missing because at present it isn't possible to include
diamond formation in the model atmospheres (due to lack of basic input data).

A number of impurities have been identified in the presolar
diamonds, including the noble gases (He, Ne, Ar, Kr, and Xe), Ba and Sr
(which are slightly enriched in r-process isotopes; Lewis et al.\ 1991),
H with $^1$H/$^2$D = 5193 (Virag et al.\ 1989;
($^1$H/$^2$D)$_{\rm terrestial}$ = 6667)
and N with $^{14}$N/$^{15}$N = 406 (Russel et al.\ 1991; 
($^{14}$N/$^{15}$N)$_{\rm terrestial}$ = 272).
The most important of these, Xe, was actually known from 
stepwise heating techniques before the grains themselves were identified
as diamonds. The Xe in the diamonds has a significant overabundance 
(compared to the solar isotopic ratios)
of the very heavy isotopes (Xe-H $\sim$ isotopes $^{134}$Xe and $^{136}$Xe) 
as well as the very light isotopes (Xe-L $\sim$ $^{124}$Xe, $^{126}$Xe).
This composition is often called Xe-HL to indicate that there is an
excess of both heavy (H) and light (L) isotopes.
There are no astronomical objects known 
(neither from observations nor from standard theories)
which have both solar $^{12}$C/$^{13}$C ratio and Xe-HL.
An explanation therefore needs to involve either
a non-standard model, not yet observationally verified, or
an assumption of the diamonds being a mixture of populations from
several different sources.

Heavy and light Xe isotopes are produced in supernovae (SN), and Clayton (1989)
therefore proposed that the meteoritic diamond grains were formed in a 
supernova that also produced the Xe-HL measured in the diamonds. 
Since the progenitor star of a supernova in the standard theories 
has an oxygen-rich atmosphere (i.e., cannot produce carbon-rich grains) and
a pure $^{12}$C interior shell (i.e., can only produce grains with
$^{13}$C/$^{12}$C $\approx$ 0), a non-standard theory was necessary.
In the extension of the model, Clayton et al.\ (1995) proposed a non-standard
SN where mixing from a $^{13}$C-rich shell occurs in the right amount to give
$^{12}$C/$^{13}$C $\approx$ 90. A non-standard
r-process was assumed too, in order to avoid the production of $^{129}$I
which decays to $^{129}$Xe and which therefore would cause a very large excess
of $^{129}$Xe, not observed in the meteorites. For a recent review of the 
standard r-, and s- neutron capture processes, see K{\"a}ppeler et al.\
(1989). Furthermore, regular r-process cannot in itself produce the very large
excess of
$^{136}$Xe characteristic for the Xe-HL measured in the presolar diamonds.
Ott (1996), however, proposed that the standard r-process is active, but that
a separation of xenon from iodine and tellurium precursors takes place in the
SN on a time scale of few hours after termination of the neutron burst in the 
SN. Since $^{136}$Xe is formed minutes after the neutron burst, and the other 
r-process Xe isotopes are formed hours ($^{134}$Xe), days ($^{131}$Xe) or
even years ($^{129}$Xe) later, a sufficiently 
early separation would allow almost infinite
amounts of $^{136}$Xe relative to the other Xe isotopes which are produced.
If a separation in the SN gas takes place 
two hours after the neutron burst, the meteoritic
$^{136}$Xe/$^{134}$Xe ratio is established in the gas, and with a small amount
of later mixing, the observed meteoritic Xe-H can be obtained. 

Detailed supernova models supporting these isotopic arguments are missing 
(as are simulations justifying, for example, the amount of $^{13}$C mixing
or why only Xe from the separated gas is implanted in the grains when they
form years after the neutron burst, etc), but
the success of fitting modified SN scenarii to the observed Xe-H
makes it likely that part of the diamond grains originates in a
supernova. 
However, there are several reasons why the bulk of the diamonds are unlikely 
to have formed in supernova: (1) The hydro-dynamical time scale of 
supernova is short compared to the time scale for carbon grain formation 
(Sedlmayr 1994). (2) The mass loss is much stronger in carbon stars with 
high C/O ratio (where pure carbon grains will form) than in carbon stars
with lower C/O ratio. If the SiC is formed in carbon stars (see next section),
there will therefore have been expelled much more diamond dust from carbon
stars (or other pure carbon grains, which are, however, not seen) into
interstellar space than SiC. (3) Carbon stars were very abundant in the
Galaxy prior to the solar system formation (due to their metallicity
dependence). The resemblance of the carbon star
$^{12}$C/$^{13}$C to the solar $^{12}$C/$^{13}$C is naturally explained
if carbon stars were the source of the solar carbon (incl.\ carbon grains),
whereas standard SN not will produce this $^{12}$C/$^{13}$C ratio.

The typical amount of Xe gas inclusion in the diamonds is 
$\approx$ 10$^{-6}$\,cm$^3$ per gram of diamond. This corresponds to 
of the order of one $^{132}$Xe atom and one $^{129}$Xe atom per 10$^7$ 
diamonds, a bit less of the isotopes $^{131}$Xe, $^{134}$Xe, and $^{136}$Xe, 
a tenth this amount of isotopes $^{128}$Xe and $^{130}$Xe, and only traces of 
$^{124}$Xe and $^{126}$Xe. A large number of 
diamonds is therefore necessary in order to perform an isotopic analysis
(in practice $\approx$ 10$^{10}$, Huss \& Lewis 1994a), and 
attempts to separate the diamonds in groups of different origin have
so far not been successful (Huss \& Lewis 1994b).
If we assume that the trapping efficiency for Xe in the 
possible population of diamonds which originated in supernovae is sufficiently
large compared to the trapping efficiency in the carbon star diamonds
(maybe because of the higher turbulent gas velocities in SN, higher densities,
etc), then the Xe-H can be explained from being connected with a 
small fraction of diamonds of pure $^{12}$C (which originated in SN), 
without altering the necessary bulk 
$^{12}$C/$^{13}$C ratio of the carbon star diamonds. We therefore propose
that the bulk of the presolar meteoritic diamonds originates in
evolved carbon stars (as in our original theory) and are mixed with a smaller 
population (from SN{\sc ii}) which has a relatively high Xe content and is 
rich in the heavy isotopes.
The content of light isotopes of Xe is very small (less than 1 atom per
10$^9$ diamonds), and can be explained as coming from SN{\sc i} (Lambert 1992)
in binary systems where the low mass component is an evolved carbon star
as in our original theory (J{\o}rgensen 1988), or as a by-process of the 
Xe production in SN{\sc ii} (Ott 1996).

\section{Silicon Carbide}

SiC is much less abundant (6\,ppm) than diamonds (1000\,ppm), but some
of the SiC grains are large enough that isotopic ratios of the Si and C
(and the abundant impurities N, Mg-Al, Ti, Ca, He, Ne) can be measured in 
individual grains (Hoppe et al.\ 1994, Lewis et al.\ 1994, Anders \& Zinner 
1993 and references therein). The understanding of their stellar origin is
therefore much better than in the case of the diamonds.

The SiC grain sizes have a large variety from less than 0.05 to 20\,$\mu$m in 
equivalent spherical diameter, with about 95\% (by mass) of the grains 
being between 0.3 and 3\,$\mu$m (Amari et al.\ 1994). 
Ion micro-probe measurements can be performed on individual grains larger than
1\,$\mu$m, and the results have made it possible to identify 
multiple stellar sources as their origin. The detailed match to the 
elemental and isotopic conditions in the He burning shell of AGB stars
(Gallino et al.\ 1990) has made it generally believed that the bulk of the 
SiC originated in carbon stars.

To be able to distinguish between various stellar origins Hoppe et al.\ (1994)
have divided the coarse (2.1--5.9 $\mu$m) SiC grains into five subgroups.
\begin{enumerate}
\item The ``mainstream'' grains have 20 $<$ $^{12}$C/$^{13}$C $<$ 120 and\\ 
   200 $<$ $^{14}$N/$^{15}$N $<$ 10\,000.
\item Grains A have $^{12}$C/$^{13}$C $<$ 3.5.
\item Grains B have 3.5 $<$ $^{12}$C/$^{13}$C $<$ 10.
\item Grains X have isotopically heavy N (13 $<$ $^{14}$N/$^{15}$N $<$ 180).
\item Grains Y have isotopically light C (150 $<$ $^{12}$C/$^{13}$C $<$ 260).
\end{enumerate}

The mainstream, type A, and type B grains have comparable patterns of Si 
isotopes, distinctly different from type X grains and from type Y grains.
The mainstream grains constitute $\approx$94\% of all coarse-grained SiC, 
whereas grains from the groups A, B, X and Y account for only 2\%, 2.5\%,
1\% and 1\% respectively.
Based on this grouping, Amari et al.\ (1995a) find that grains X could 
originate from a supernova (SN{\sc ii}), and Lodders \&
Fegley (1995) find that grains A and B can be at least qualitatively
understood if they originate from J-type carbon stars or carbon stars that
have not experienced much dredge-up of He-shell material.

It is seen that the isotopic variations among the grains are very large.
$^{12}$C/$^{13}$C varies by a factor more than 350, 
$^{14}$N/$^{15}$N varies by 300 times
(and $^{30}$Si/$^{28}$Si by a factor of 3).
Variations in noble gases associated with SiC are large as well (Ott 1993).
The two noble gas components, s-process Xe and Neon-E (i.e.,
essentially pure $^{22}$Ne), show opposite correlations 
with grain size, s-process Xe being most abundant in fine-grained SiC and 
Ne-E in coarse-grained SiC (Lewis et al.\ 1994). Other elements which show 
s-process indications include s-process Kr, Ba, Sr, Ca, Ti, Nd and Sm.

The proportions of $^{80,86}$Kr vary with release temperature of the gas.
This variation reflects branching of the s-process at the radioactive
progenitors,
$^{79}$Se and $^{85}$Kr (Ott et al.\ 1988).  These branchings depend
sensitively on neutron density and temperatures in the s-process region, and
the $^{80,86}$Kr can therefore provide clues about in which stars
the SiC formed, or if the stellar type is already known it can 
put constraints on the detailed modelling of these stars.

\section{Graphite}

The presolar graphite
isolated from meteorites lies at the graphitic end of the continuum between
kerogen, amorphous carbon, and graphite. It is not very abundant (less than 
1\,ppm), and it is much more complicated to extract 
than SiC and micro-diamonds (Amari et al.\ 1994). 
Presolar graphite occurs solely in the form of spherules, $0.8 - 8 \mu$m
in diameter, while graphite grains of other sizes and shapes have normal 
composition and are believed to have been formed in the solar nebula
(Zinner et al.\ 1990).  
The presolar graphite has a very broad $^{12}$C/$^{13}$C distribution,
with $^{12}$C/$^{13}$C ratios ranging from 7 to 4500,
whereas the $^{14}$N/$^{15}$N ratios range from 193 to 680 
(Zinner et al.\ 1995).

The noble gases show systematic trends with sample density, suggesting
more than one kind of graphite.  Some have almost
mono-isotopic $^{22}$Ne. Others contain neon with a somewhat higher
$^{20}$Ne/$^{22}$Ne ratio and are accompanied by s-process Kr, $^{4}$He and
other noble gases (Amari et al.\ 1995b).

The carbon and nitrogen isotopic ratios found in the grains indicate that 
they come from stellar
sources dominated by H-burning rather than from sources dominated by He-burning
(Amari et al.\ 1993). H-burning in the CNO cycle
produces isotopically heavy carbon ($^{13}$C) and 
light nitrogen ($^{14}$N), in qualitative agreement with the measurements
(Zinner et al.\ 1989; Hoppe et al.\ 1994). Systematic measurements 
of isotopic ratios of several other elements were recently done by
Hoppe et al.\ (1995).

\end{document}